\documentclass[a4paper,11pt]{article}
\pdfoutput=1
\usepackage{jcappub}
\usepackage{graphicx}
\usepackage{amssymb}
\usepackage{color}
\usepackage[normalem]{ulem}

\newcommand{\Neff}{\ensuremath{N_{\rm eff}}}
\newcommand{\lcdm}{$\Lambda$CDM}
\newcommand{\kanu}{\ensuremath{\kappa_\nu}}
\newcommand{\lcdmk}{$\Lambda$CDM$+\kanu$}
\newcommand{\lcdmm}{$\Lambda$CDM$+m_\nu$}
\newcommand{\lcdmkm}{$\Lambda$CDM$+\kanu+m_\nu$}

\title{\boldmath Cosmological bounds on neutrino statistics}
 
\author[a]{P.F.\ de Salas,}
\author[a]{S.\ Gariazzo,}
\author[b]{M.\ Laveder,}
\author[a]{S.\ Pastor,}
\author[c]{O.\ Pisanti}
\author[d,e]{and N.\ Truong}

\affiliation[a]{Instituto de F\'{\i}sica Corpuscular
(CSIC-Universitat de Val\`{e}ncia)\\ 
Parc Cient\'{\i}fic UV, C/ Catedr\'atico Jos\'e Beltr\'an, 2\\ E-46980 Paterna (Valencia), Spain}

\affiliation[b]{Dipartimento di Fisica e Astronomia ``G. Galilei'', Universit\`a di Padova,\\
and
INFN, Sezione di Padova,
Via F. Marzolo 8, I--35131 Padova, Italy}

\affiliation[c]{Dipartimento di Fisica E. Pancini, Universit\`a di Napoli Federico II,\\
and INFN, Sezione di Napoli, Via Cintia, I-80126 Napoli, Italy.}

\affiliation[d]{MTA-E\"otv\"os University Lend\"ulet Hot Universe Research Group\\
P\'azm\'any P\'eter s\'et\'any 1/A, Budapest, 1117, Hungary}
\affiliation[e]{Institute For Interdisciplinary Research in Science and Education \\
ICISE, Ghenh Rang, 590000 Quy Nhon, Vietnam}

\emailAdd{pabferde@ific.uv.es}
\emailAdd{gariazzo@ific.uv.es}
\emailAdd{marco.laveder@pd.infn.it}
\emailAdd{pastor@ific.uv.es}
\emailAdd{pisanti@na.infn.it}
\emailAdd{truongnhut@caesar.elte.hu}

\abstract{
We consider the phenomenological implications of the violation of the Pauli exclusion principle for neutrinos, focusing on cosmological observables such 
as the spectrum of Cosmic Microwave Background anisotropies, Baryon Acoustic Oscillations and the primordial abundances of light elements. Neutrinos that behave (at least partly) as bosonic particles have a modified equilibrium distribution function that implies a different influence on the evolution of the Universe that, in the case of massive neutrinos, can not be simply parametrized by a change in the effective number of neutrinos. Our results show that, despite the precision of the available cosmological data, only very weak bounds can be obtained on neutrino statistics, disfavouring a more {\em bosonic} behaviour at less than $2\sigma$.
}
\begin{document}
\maketitle
\flushbottom

\section{Introduction} 
\label{sec:intro}

Neutrinos are probably the most peculiar particles of the Standard Model. Due to the weakness of their interactions, 
the experimental study of neutrinos has been only developed in the last decades, leading to unexpected
properties such as neutrino mixing. Nowadays, thanks to a plethora of experimental results on the measurement of reactor,
accelerator, atmospheric and solar neutrinos, the phenomenon of flavour oscillations is well understood (see e.g.\ \cite{deSalas:2017kay}).
However, terrestrial experiments are not always the best option for constraining some neutrino properties and alternative astroparticle probes
could be employed instead. For instance, by studying the relics of the Big Bang,
such as the Cosmic Microwave Background (CMB) radiation
and the distribution of matter at large scales, one can gain
knowledge on elementary particles, even those with extremely weak interactions like neutrinos,
which played a very important collective role in the Universe evolution \cite{NuCosmo}.
There have been many cosmological studies on various neutrino properties, as for example
their total mass, 
the possible existence of light sterile neutrinos, their contribution to the so-called dark 
radiation (via the effective number of neutrino species, $N_{\rm eff}$),
or neutrino self-interactions (see e.g.\ \cite{Lesgourgues:2014zoa,Lattanzi:2017ubx} for a review).

In this paper, our aim is to investigate the possible cosmological bounds on another property of these tiny particles: the neutrino statistics.
From the theoretical point of view, this issue was addressed in an original work by W.\ Pauli \cite{Pauli:1940zz},
which results in the well-known spin-statistics theorem. According to it, the wave function of integer-spin particles should be symmetric, 
meaning that it is invariant under permutations of the position of identical particles.
Those particles are categorised as bosons, and an ensemble of bosons in thermal equilibrium obeys the {\it Bose-Einstein (BE)} distribution.
On the other hand, particles with half-integer spin should be represented by antisymmetric wave functions, which change sign under position permutations.
These particles are named fermions, and their thermal distribution is the {\it Fermi-Dirac (FD)} distribution.
This is the case, for instance, of neutral leptons such as neutrinos. However, contrary to the case of electrons and nucleons, a 
possible violation of the Pauli exclusion principle for neutrinos is not yet experimentally excluded, although it affects
elementary processes which involve identical neutrinos, such as double beta decay \cite{Dolgov:2005qi,Barabash:2007gb}.
Such a violation was discussed in a series of theoretical papers, although no satisfactory model has been proposed so far 
(more discussion in \cite{Dolgov:2005qi} and references therein).

Here we follow previous works and take a purely phenomenological approach for neutrino statistics, described by a single continuous parameter $\kanu$
which describes purely fermionic ($\kanu=+1$) and purely bosonic ($\kanu=-1$) neutrinos.
Previous analyses 
have described the main effects of $\kanu\neq 1$ on the early Universe, in particular on
Big Bang Nucleosynthesis (BBN) \cite{Cucurull:1995bx,Dolgov:2005qi,Dolgov:2005mi}, including a change in the contribution to the relativistic degrees of freedom 
(see also \cite{Iizuka:2014wma,Iizuka:2016flh,Iizuka:2017zff}). A modified neutrino statistics would also lead to a different behaviour as 
hot dark matter \cite{Hannestad:2005bt,Brandbyge:2017tdc} and, in the extreme case of bosonic relic neutrinos, even as light
as axions, they could condense and act as the cosmological dark matter \cite{Dolgov:2005qi}. Other astrophysical consequence
of the violation of the Pauli principle would be a different influence on the dynamics of the supernova collapse
\cite{Dolgov:2005qi}, as well as on a future detection of supernova neutrinos \cite{Choubey:2005vi}.

Our paper extends previous works on the cosmological implications of a modified neutrino statistics, studying 
the bounds on $\kanu$ that can be obtained from the precise cosmological data currently available, 
both from BBN and the combined analysis of CMB measurements 
and Baryon Acoustic Oscillations (BAO) data, including for the first time neutrino mass as a free parameter.
In addition, we want to check
how the currently known cosmological tensions
are affected by
a deviation from the purely fermionic nature of neutrinos.
For instance, CMB measurements from the Planck satellite \cite{Ade:2015xua,Aghanim:2016yuo} appear to prefer a significantly lower value of the current 
expansion rate
$H_0 = 100\, h \, \mathrm{km}\,\mathrm{s}^{-1}\,\mathrm{Mpc}^{-1}$
than what was derived from the type Ia Supernovae Survey \cite{Riess:2016jrr}.
Another tension that involves the same Planck mission affects matter fluctuations at small scales, 
quantified by the parameter $\sigma_8$ (a measure of the mean matter fluctuations in a sphere with a radius of 8~Mpc).
Recent determinations of this parameter, using KiDS-450 \cite{Hildebrandt:2016iqg}
and Dark Energy Survey (DES) \cite{Abbott:2017wau} data, result in a smaller
value than the one obtained by Planck, exhibiting a discrepancy at a level of a bit more than 2$\sigma$ in the pure \lcdm\ model.
Finally, concerning BBN, we have the well-known discrepancy between the theoretical value of the primordial $^7$Li abundance and its experimental determination 
from the Spite plateau \cite{Spite:1982dd}, the so-called {\it $^7$Li problem}. 
While these tensions may arise from not-yet-revealed observational systematics, they can also be viewed as signals that open doors to new physics.
  
This work is organised as follows.
In section \ref{sec:stats} we describe the main phenomenological effects of a modified neutrino statistics on the cosmological evolution.
The analysis of the corresponding cosmological model taking into account present data, from CMB alone and in combination with BAO measurements,
is presented in section \ref{sec:cosmo}, while the effects and bounds from primordial nucleosynthesis are discussed  
in section \ref{sec:bbn}. Finally, we summarize our main results and present our conclusions.

\section{Cosmological effects of a modified neutrino statistics}
\label{sec:stats}

In our phenomenological study we consider that the energy distribution of neutrinos in a thermal bath,
using natural units $c=\hbar=k_{\rm B}=1$, is given by
\begin{equation}
f_\nu^{\rm eq}(E)=\frac{1}{e^{E/T_\nu}+\kanu}~,
\label{eq1}
\end{equation}
where $T_\nu$ and $E$ are, respectively, the neutrino temperature and energy,
and $\kanu$ is the {\em Fermi-Bose parameter} that characterizes the neutrino statistics,
which can vary from $-1$ to $+1$.
Such a parametrization is motivated by earlier works
on possible deviations of the neutrino statistics from the Fermi-Dirac distribution
\cite{Dolgov:2005qi,Dolgov:2005mi,Barabash:2007gb}.
The strongest bound on the parameter $\kanu$ up to now
comes from the observation of two-neutrino double beta decay processes, which 
leads to the lower bound $\kanu>-0.2$~\cite{Barabash:2007gb,Barabash:2009jx}\footnote{This bound could be improved soon
from a careful analysis of NEMO-3 data (A.S.\ Barabash, private communication).}. Thus,
a 100\% violation of Fermi statistics for neutrinos is disfavoured, but a large admixture of the bosonic component
can still be present.

The parameter $\kanu$ should be also present in the statistical factor of the collision integral of any process involving neutrinos. For instance,
in the case of the elastic scattering $\nu_1+l_1 \leftrightarrow \nu_2+l_2$ the corresponding statistical factor would be
\begin{equation}
F = f_\nu (k_1) f_l(p_1)[1 - f_l(p_2)] [1 - \kanu f_\nu(k_2)] - f_\nu(k_2)f_l(p_2)[1 - f_l(p_1)] [1- \kanu f_\nu(k_1)]~.
\end{equation}
The generalized form $f_\nu^{\rm eq}(E)$ in eq.\ \eqref{eq1} guarantees that, in thermal equilibrium, the collision integral of the process vanishes 
also for the case of mixed neutrino statistics.

In the early Universe, neutrinos decouple from the rest of the primeval plasma at a temperature around 1~MeV, when they still behave as ultrarelativistic particles \cite{NuCosmo}.
Afterwards, their distribution function
will be described by the equilibrium distribution function
of a massless fermion in the standard case or by the form in eq.\ \eqref{eq1} with the substitution $E(p)\to p$ in the case of a mixed neutrino statistics.

\begin{figure*}[t]
\centering
\includegraphics[width=0.8\textwidth]{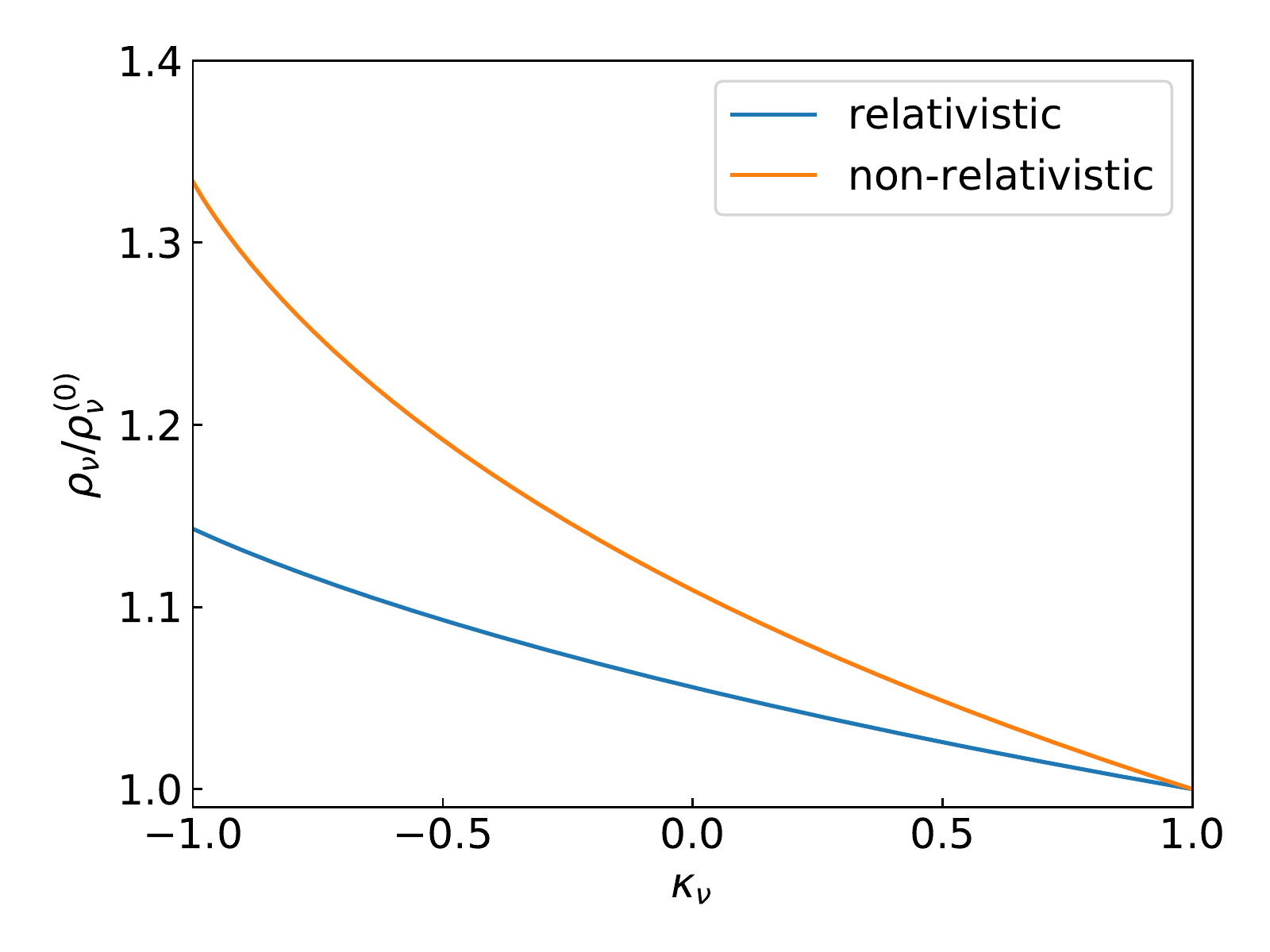}
\caption{\label{fig:rhonu}
Ratio of neutrino energy densities $\rho_\nu/\rho^{(0)}_\nu$ as a function of $\kanu$, where $\rho^{(0)}_\nu$ corresponds to the purely fermionic case.}
\end{figure*}

The first and obvious effect of $\kanu\neq 1$ is a change in the contribution of neutrinos to the cosmological energy density, given by
\begin{equation}
\rho_\nu=
\frac{g_\nu}{2\pi^2}
\int_0^\infty dp\, p^2 E(p)\times f_{\nu}(p)
=\frac{g_\nu}{2\pi^2}
\int_0^\infty dp\frac{p^2 E(p)}{\exp(p/T_\nu)+\kanu}\qquad,
\label{eq2}
\end{equation}
where $g_\nu=2$ specifies the neutrino degrees of freedom per state.
Therefore, varying the neutrino statistics has an impact on the neutrino energy density throughout the entire cosmic history.

For massless neutrinos, their modified energy density can be increased up to a factor 8/7 with respect to its value for fermionic neutrinos,
as can be seen in Fig.~\ref{fig:rhonu} for different values of $\kanu$.
This factor arises from the calculation of the integral
in eq.\ \eqref{eq2} for bosonic instead of fermionic particles.
The modification in the energy density is fully equivalent 
to a change in the effective number of neutrinos, $\Neff$, the parameter that 
describes the impact of relativistic particles other than photons on the total energy density of radiation.
For three neutrino states, this amounts to a larger
$\Neff$, with a maximum difference with respect to the standard value
$\Neff^{\rm std}=3.045$ \cite{deSalas:2016ztq}
of $\Delta \Neff \equiv \Neff-\Neff^{\rm std} \simeq 0.43$
for purely bosonic neutrinos.
The variation in $\Neff$ is one of the required changes for a proper calculation of the outcome of primordial nucleosynthesis, together with the
inclusion of the spectrum in eq.\ \eqref{eq1} for electron neutrinos and antineutrinos in the weak processes that relate neutrons and protons, 
as we discuss in section \ref{sec:bbn}.

For the more realistic case of massive neutrinos, the effect of a mixed statistics to the energy density is different at early and current times.
At the epochs when relic neutrinos are still relativistic, their contribution to $\rho_\nu$
can be parametrized with an enhanced $\Neff$, as we just described.
However, after their non-relativistic transition the neutrino energy in eq.\ \eqref{eq2}
is replaced by the neutrino mass.
In this case, the ratio 
of the energy densities $\rho_\nu(\kanu)/\rho^{(0)}_\nu$ also
grows for smaller values of $\kanu$, but now reaching a maximum of 4/3 for purely bosonic neutrinos, as depicted in Fig.~\ref{fig:rhonu}.

\begin{figure*}[t]
\centering
\includegraphics[width=0.49\textwidth]{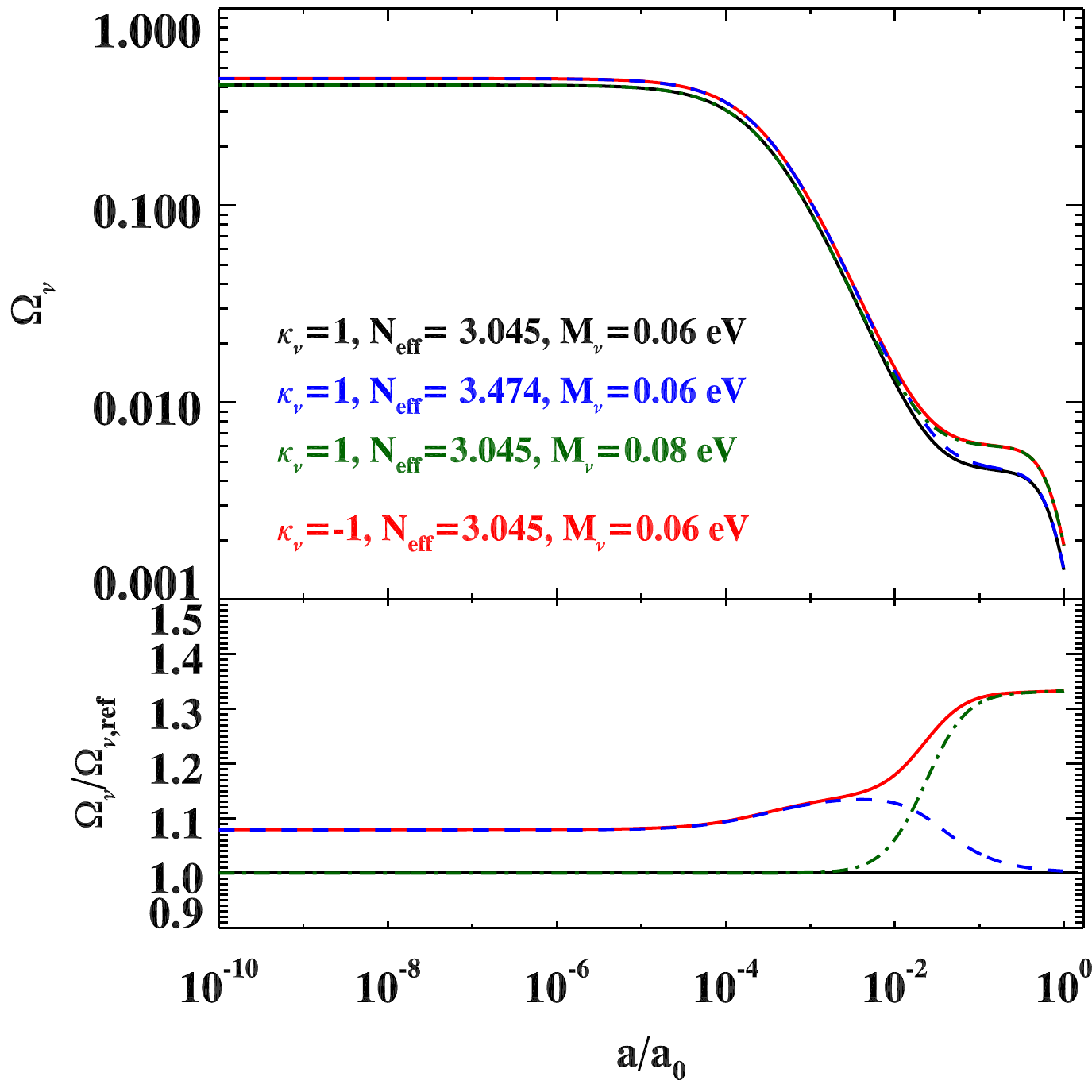}
\includegraphics[width=0.49\textwidth]{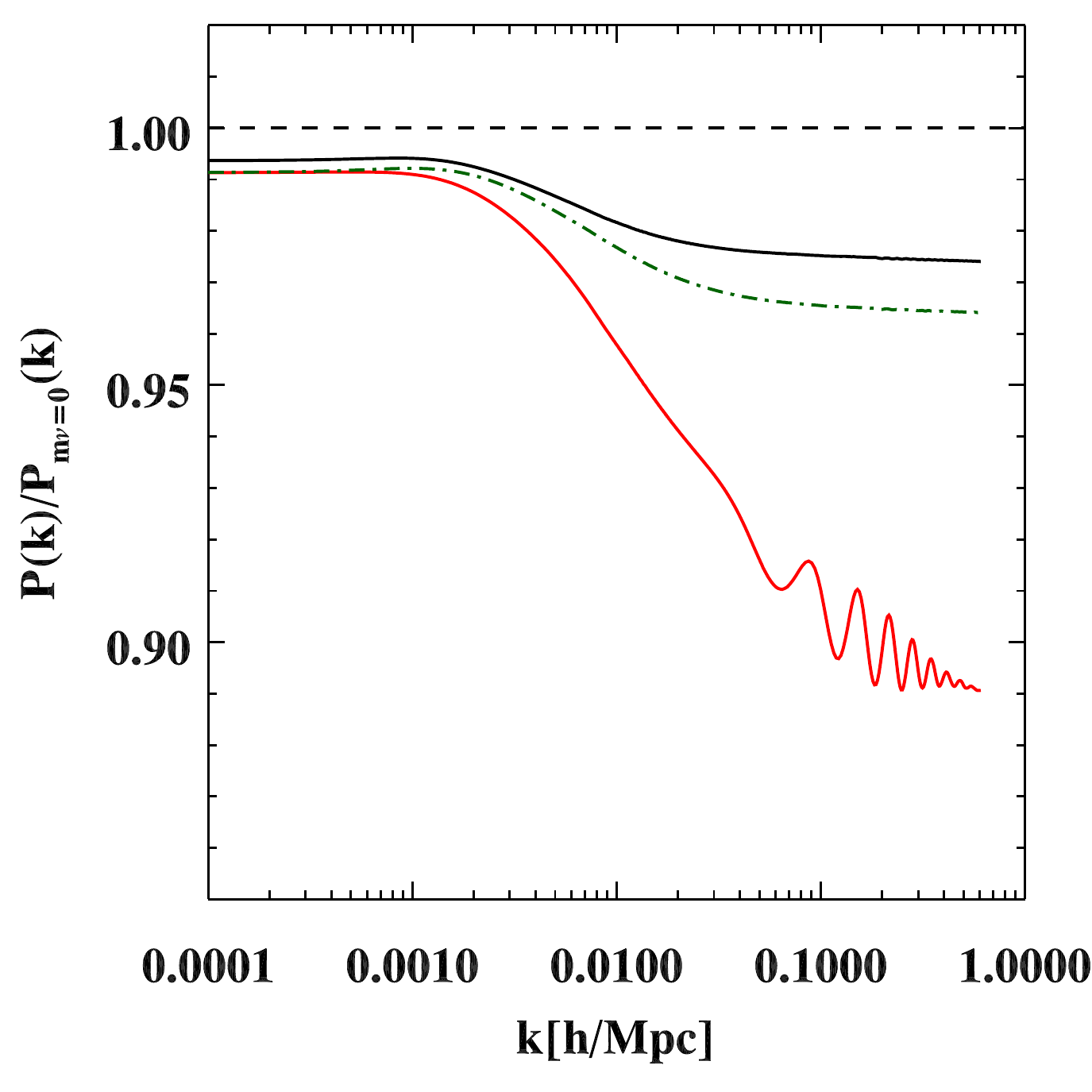}
\caption{\label{fig:1}
{\it Left panel:} Neutrino energy density
in units of the critical energy density of the Universe, $\Omega_\nu=\rho_\nu/\rho_{\rm tot}$,
as a function of the scale factor $a$,
for different values of $\kanu$, $\Neff$,
and the sum of the neutrino masses $M_\nu$.
$\Omega_{\nu,\text{ref}}$ is the neutrino energy density fraction
for $\Neff=3.045$ and $M_\nu=\sum m_\nu=0.06$~eV.
{\it Right panel:} Degree of late-time matter power suppression for different values of $\kanu$ and neutrino mass.
Parameter values are the same for curves with the same colour in
the two panels.
}
\end{figure*}

Therefore, the full effect of having a modified neutrino statistics on
the cosmological energy density is not equivalent to a global rescale of
\Neff\ or of the neutrino masses.
In the left panel of Fig.~\ref{fig:1} we illustrate how the neutrino energy density changes
during the cosmological evolution for fermionic (black) and bosonic (red) neutrinos,
in comparison to the standard case of $\kanu=1$,
when $\Neff$ is manually increased by 0.43 (dashed blue)
and when the neutrino mass is increased by a factor of $4/3$
(dashed-dotted green).
As we can see, for a lower value of $\kanu$
both the early- and late-time neutrino energy densities
are increased as expected from eq.~\eqref{eq2}.
The case of FD neutrinos with an increased \Neff{}
is equivalent at early times to the case of BE neutrinos,
but not at late times, when neutrinos are no longer relativistic.
On the other hand, neutrinos with an increased mass mimic the effect of the purely bosonic case
only at late times, because the mass does not have any effect at the early stage.
In the right panel of Fig.~\ref{fig:1} we show the effect of the increased late-time neutrino energy
on the current matter power spectrum.
A larger late-time neutrino energy density causes
more suppression to the matter power spectrum, an effect that is similar to having an increased neutrino mass.
Therefore, if one wants to test a mixed neutrino statistics,
it is clear that the entire cosmological evolution must be computed 
taking into account the neutrino distribution with a free $\kanu$, as well as the possible values of neutrino masses. 

\section{Bounds from CMB and BAO data}
\label{sec:cosmo}

In this section we study whether cosmological observables such as CMB measurements can provide any bound on neutrino statistics.
We base our study on the standard \lcdm\ cosmological model,
parametrized by the usual six parameters:
the baryon and cold dark matter energy densities $\omega_b$ and $\omega_c$,
the optical depth to reionization $\tau$,
the ratio of the sound horizon to the angular diameter distance at decoupling $\theta$,
and the tilt and amplitude of the power spectrum of primordial curvature perturbations $n_s$ and $A_s$.
In addition, we will consider its extensions:
one with fermionic neutrinos and free neutrino masses (\lcdmm\ model),
one where we fix the neutrino masses and vary
the parameter $\kanu$ describing the neutrino statistics (\lcdmk\ model)
and the case where both $\kanu$ and $m_\nu$ are free (\lcdmkm\ model).
When considering the models with varying neutrino masses,
we adopt a degenerate case for the neutrino mass eigenstates, i.e.\ three identical masses
$m_\nu=\sum m_{i}/3$.
This is a good approximation,
since current cosmological experiments are not yet precise enough
to discriminate the three separate mass states.
When fixing the neutrino masses, we will assume $m_\nu=0.02$~eV,
corresponding to the minimum $M_\nu = \sum m_{i}=0.06$~eV allowed by neutrino oscillations
in the context of normal mass ordering.

The theoretical models are computed with a modified version of the Boltzmann code \texttt{CLASS} \cite{Blas:2011rf},
while to explore the parameter space we employ its companion code \texttt{MontePython} \cite{Audren:2012wb}
together with the \texttt{MultiNest} \cite{Feroz:2008xx,Bridges:2008ta,Feroz:2013hea}
nested sampling algorithm. Concerning the experimental measurements,
we use \textbf{CMB} data from the Planck 2015 release%
\footnote{The Planck likelihoods are publicly available
at \href{http://www.cosmos.esa.int/web/planck/pla}{www.cosmos.esa.int/web/planck/pla}.}:
the high-$\ell$ ($30 \leq \ell
\leq 2508$) and
the low-$\ell$ ($2 \leq \ell \leq 29$) temperature auto-correlation likelihoods, 
the Planck polarization likelihood
at low-multipoles ($2 \leq \ell \leq 29$)
and the \texttt{lensing} likelihood from the 4-point correlation function~\cite{Aghanim:2015xee,Adam:2015rua,Ade:2015xua,Ade:2015zua}.
We will also consider \textbf{BAO} measurements, including data from the 6DF \cite{Beutler:2011hx} at $z=0.106$,
SDSS DR7 MGS \cite{Ross:2014qpa} at $z=0.15$
and the two BOSS DR10/DR11 results \cite{Anderson:2013zyy},
from LOWZ at $z=0.32$ and from CMASS at $z=0.57$.

\begin{figure*}[t]
\centering
\includegraphics[width=0.49\textwidth]{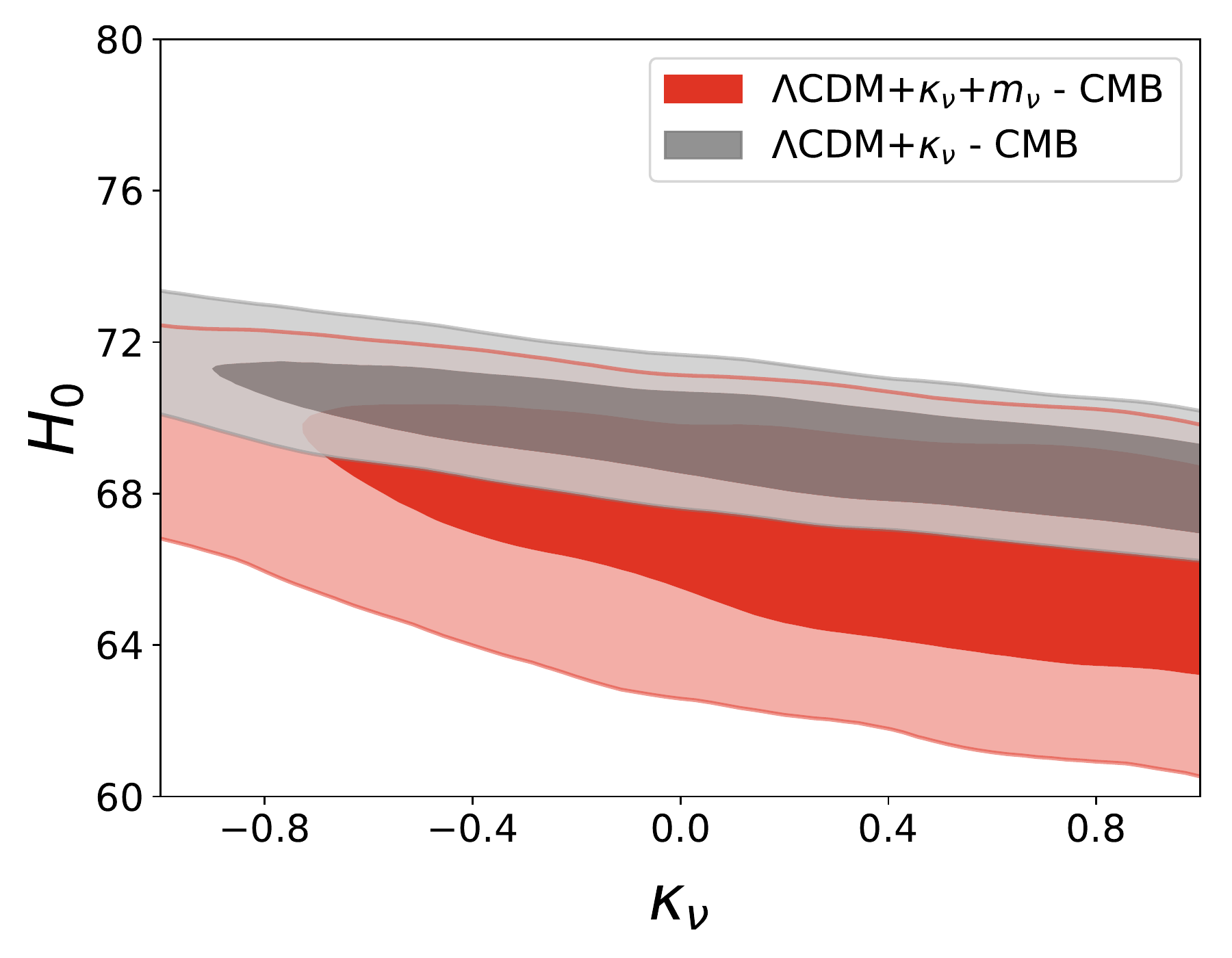}
\includegraphics[width=0.49\textwidth]{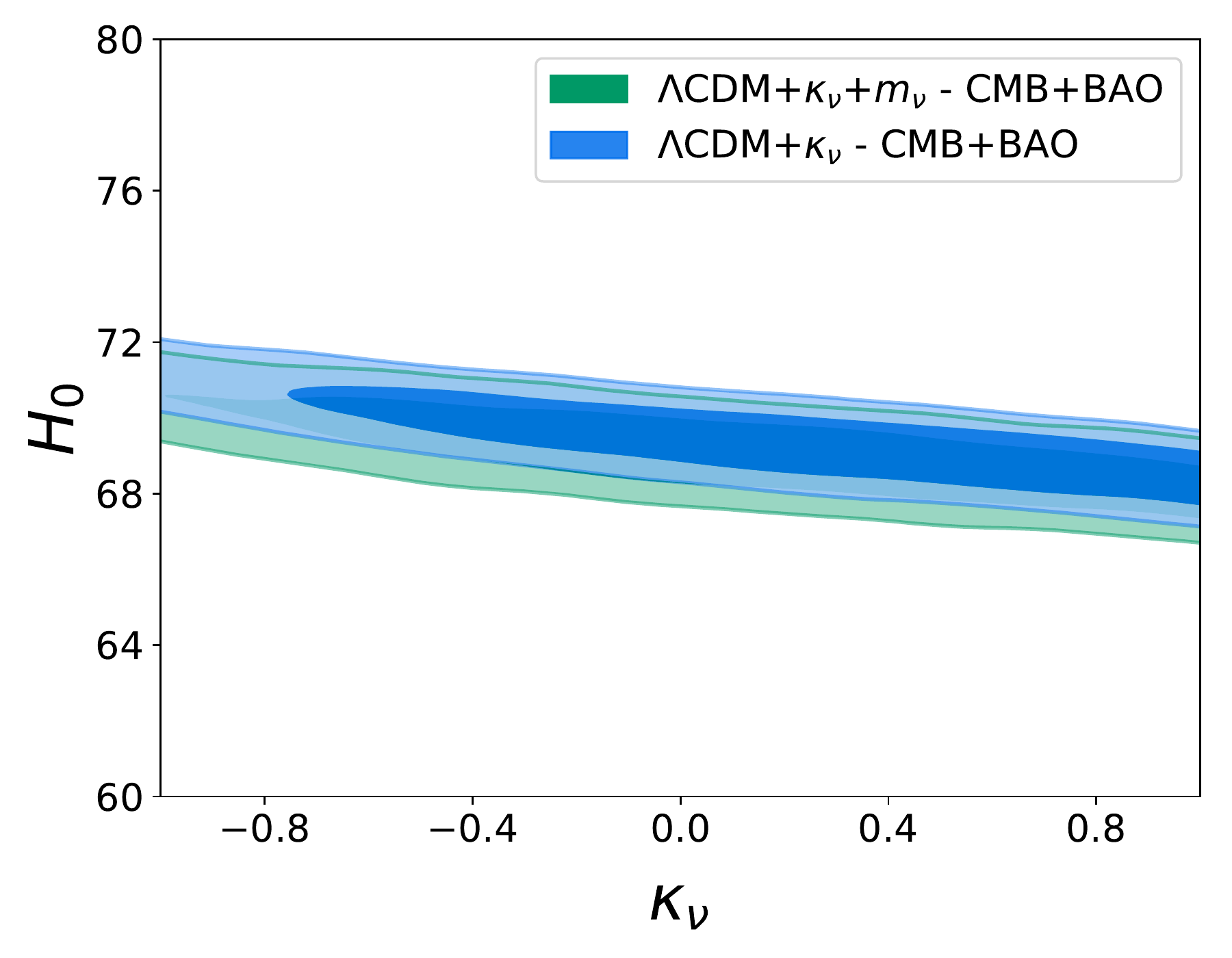}

\caption{\label{fig:k_vs_h0}
Degeneracy between $\kanu$ and $H_0$,
considering two different data combinations (CMB and CMB+BAO) and theoretical models (\lcdmk\ and \lcdmkm).
Areas represent $1\sigma$ and $2\sigma$ credible regions.}
\end{figure*}

We do not show here a full comparison
between the results obtained in
the standard \lcdm\ model and
the extended \lcdmk\ model,
in which we fix a total neutrino mass $\Sigma m_\nu = 0.06$~eV and
the standard effective number of neutrino species $\Neff=3.045$ 
for $\kanu=+1$ \cite{deSalas:2016ztq}.
The reason is that most of the cosmological parameters are affected
by the new varying quantity $\kanu$ only through
its partial degeneracy with $\Neff$, discussed before,
and the constraints do not change significantly.
For example, we show in fig.~\ref{fig:k_vs_h0} that this degeneracy
forces an expected correlation of $\kanu$ with the present value of the 
Hubble parameter $H_0$, which in turn impacts all the other parameters.
Since $m_\nu$ and $\kanu$ have some degree of degeneracy and the neutrino mass is also correlated with $H_0$,
the size of the 2D allowed contours is much larger in the \lcdmkm\ model.
On the contrary, since BAO data strongly constrain $m_\nu$,
the degeneracy is reduced when the CMB+BAO dataset is considered.

\begin{figure*}[t]
\centering
\includegraphics[width=0.6\textwidth]{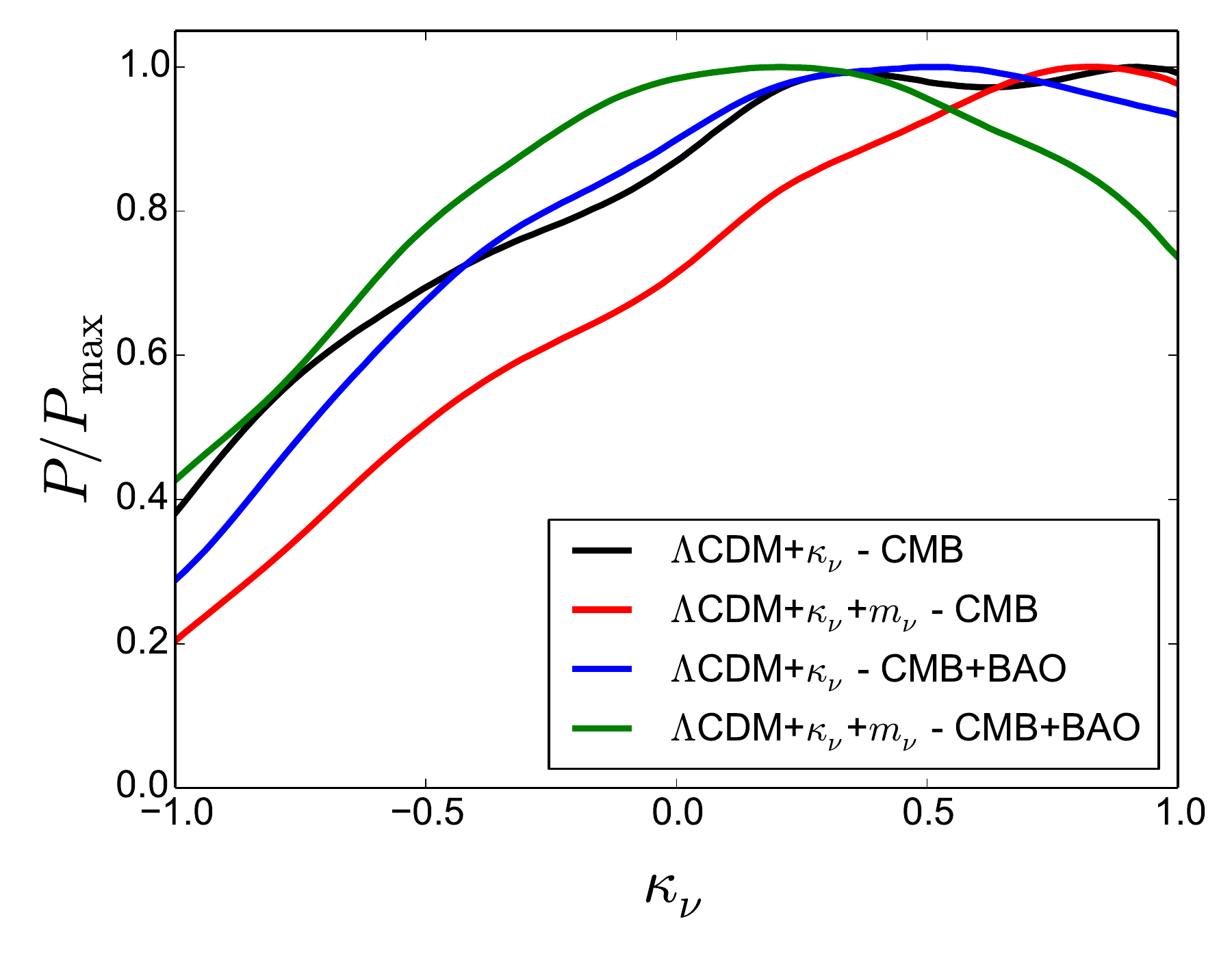}
\caption{\label{fig:kappanu}
Marginalized posterior distribution for the parameter $\kanu$
from CMB data alone and in combination with BAO,
in the \lcdmk\ and
\lcdmkm\ models. 
}
\end{figure*}

\begin{figure*}[t]
\centering
\includegraphics[width=0.7\textwidth]{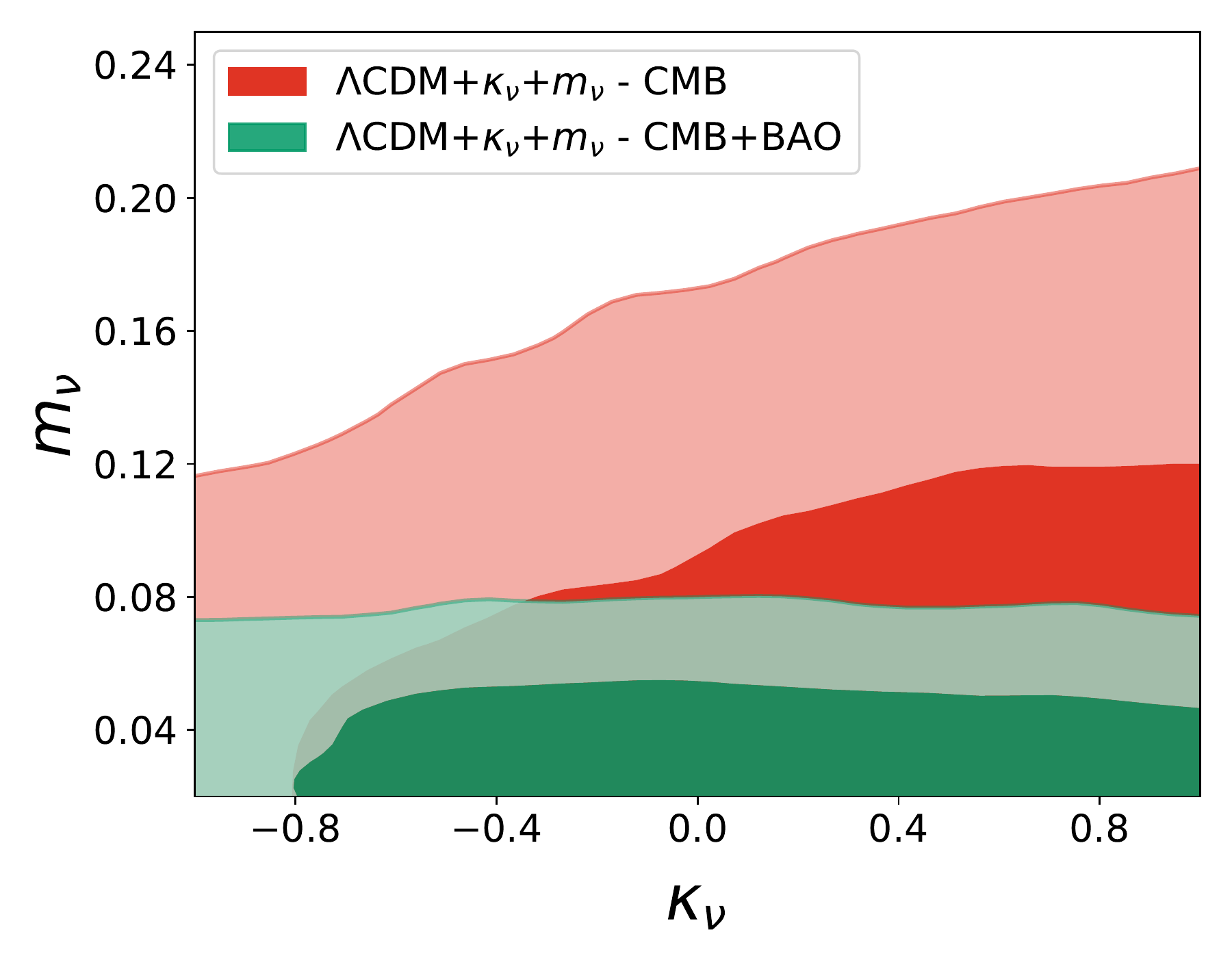}
\caption{\label{fig:k_vs_m}
Correlations between $\kanu$ and $m_\nu$
in the \lcdmkm\ model,
using CMB data alone or in combination with BAO.
Areas represent $1\sigma$ and $2\sigma$ credible regions.
}
\end{figure*}

The 1D marginalized posteriors on $\kanu$ from our analysis are presented in fig.~\ref{fig:kappanu}.
Since from the CMB point of view a change in $\kanu$
is nearly equivalent to increasing $\Neff$ by 0.43 or less,
the current experimental sensitivity is not sufficient
to obtain a strong bound on neutrino statistics: the current error on $\Neff$ is of the order of 0.2 at 68\% CL
and the maximum variation $\Delta\Neff\simeq0.43$
is within the 95\% CL bound.
The best we can obtain is then a lower limit
\begin{equation}
\kanu > -0.18\ (68\%, \text{ CMB, }\Lambda\text{CDM}+\kanu)~, 
\label{eq4}
\end{equation}
which means that the purely bosonic case is still allowed at the 95\% CL, although disfavoured at more than 68\% CL.
This result is consistent with the lower bound obtained from the analysis of 
two-neutrino double beta decay ($\kanu>-0.2$),
and implies that mixed statistics of neutrinos is still allowed given the current CMB data.
The 68\% lower limit changes only a bit
if one considers a different data combination for the same model
or an extended cosmological model with CMB data only,
\begin{eqnarray}
\kanu &>& -0.15\ (68\%, \text{ CMB+BAO, }\Lambda\text{CDM}+\kanu)~, 
\label{eq:cmbbao_lcdmkm}\\
\kanu &>& -0.06\ (68\%, \text{ CMB, }\Lambda\text{CDM}+\kanu+m_\nu)~, 
\end{eqnarray}
Instead, a difference appears in the analysis of the
\lcdmkm\ model with CMB+BAO data.
In this case, indeed,
we find a 68\% range of $-0.38<\kanu < 0.83$, that does not include the pure FD and BE distributions.
As demonstrated in ref.~\cite{Hannestad:2017ypp}, however,
we recall that a flat prior is not the best way to sample the parameter space
of a constrained parameter like $m_\nu$ or $\kanu$,
and the bounds on these parameters may be influenced by the chosen prior.
We think that this is exactly the case and
that the anomalous behaviour of the $\kanu$ bounds
in this case is driven by the fact that
there is a correlation between $m_\nu$ and $\kanu$
(see fig.~\ref{fig:k_vs_m}),
but BAO data strongly prefer values of $m_\nu$ close to the lower allowed value.
As a consequence, volume effects or shot noise problems
are more likely to appear here,
since we are dealing with an extreme of the prior range,
where the Bayesian inference may suffer more a prior dependence.
Moreover, as reported in eq.\ \eqref{eq:cmbbao_lcdmkm},
when we fix the neutrino masses to the minimum of our prior,
i.e.\ when we consider the \lcdmk\ model,
the bound is perfectly compatible with the one obtained from CMB data alone
and there is no exclusion of the FD distribution.
Another interesting thing we can learn from fig.~\ref{fig:k_vs_m}
is that there is less freedom for BE neutrinos when they have
large masses.
The reason is that the late time energy density for bosonic neutrinos
is 4/3 larger than the FD one for the same masses,
and consequently the neutrino mass is more constrained
for $\kanu\simeq-1$.

\begin{figure*}
\centering
\includegraphics[width=0.55\textwidth]{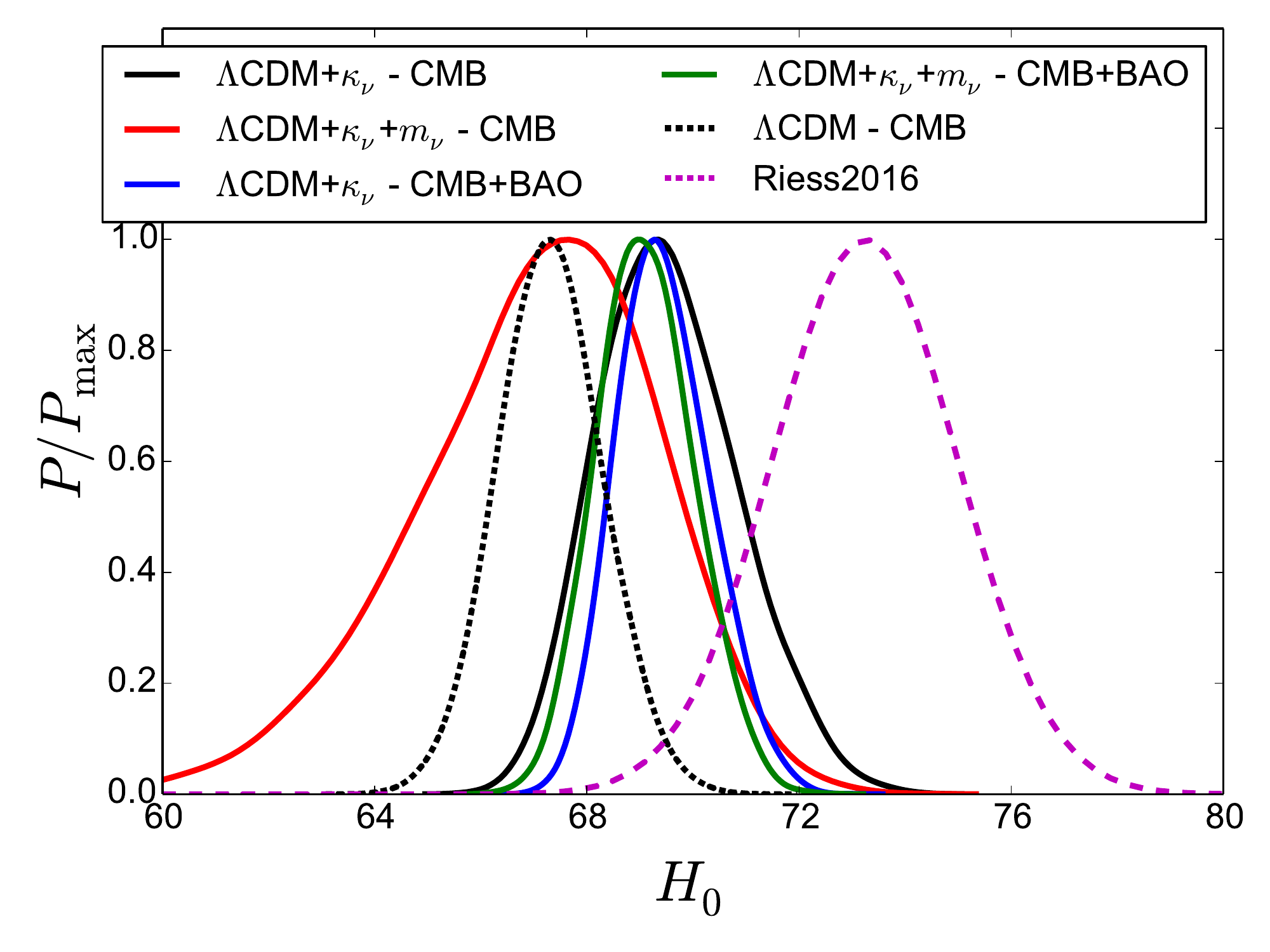}
\includegraphics[width=0.55\textwidth]{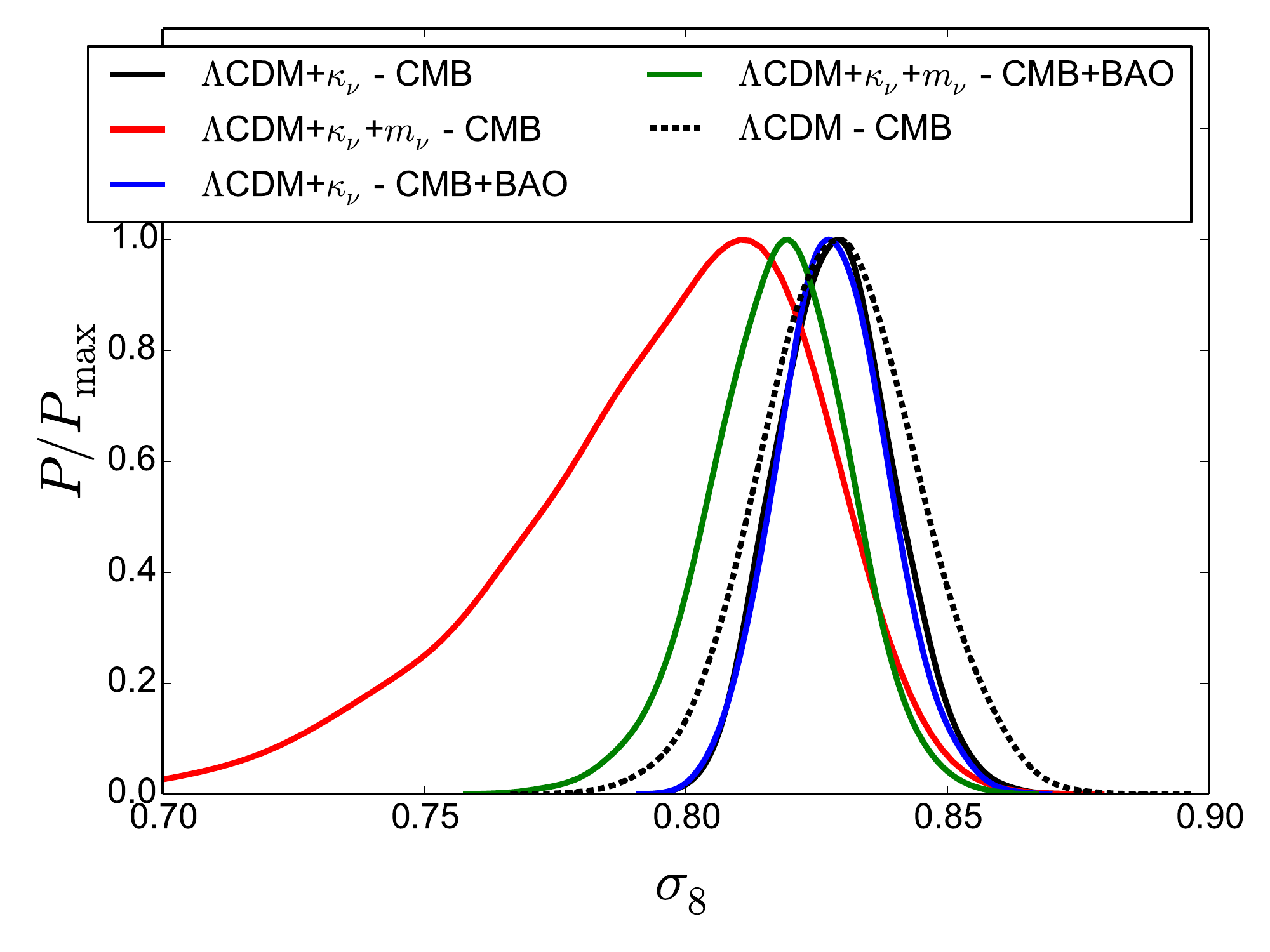}
\includegraphics[width=0.55\textwidth]{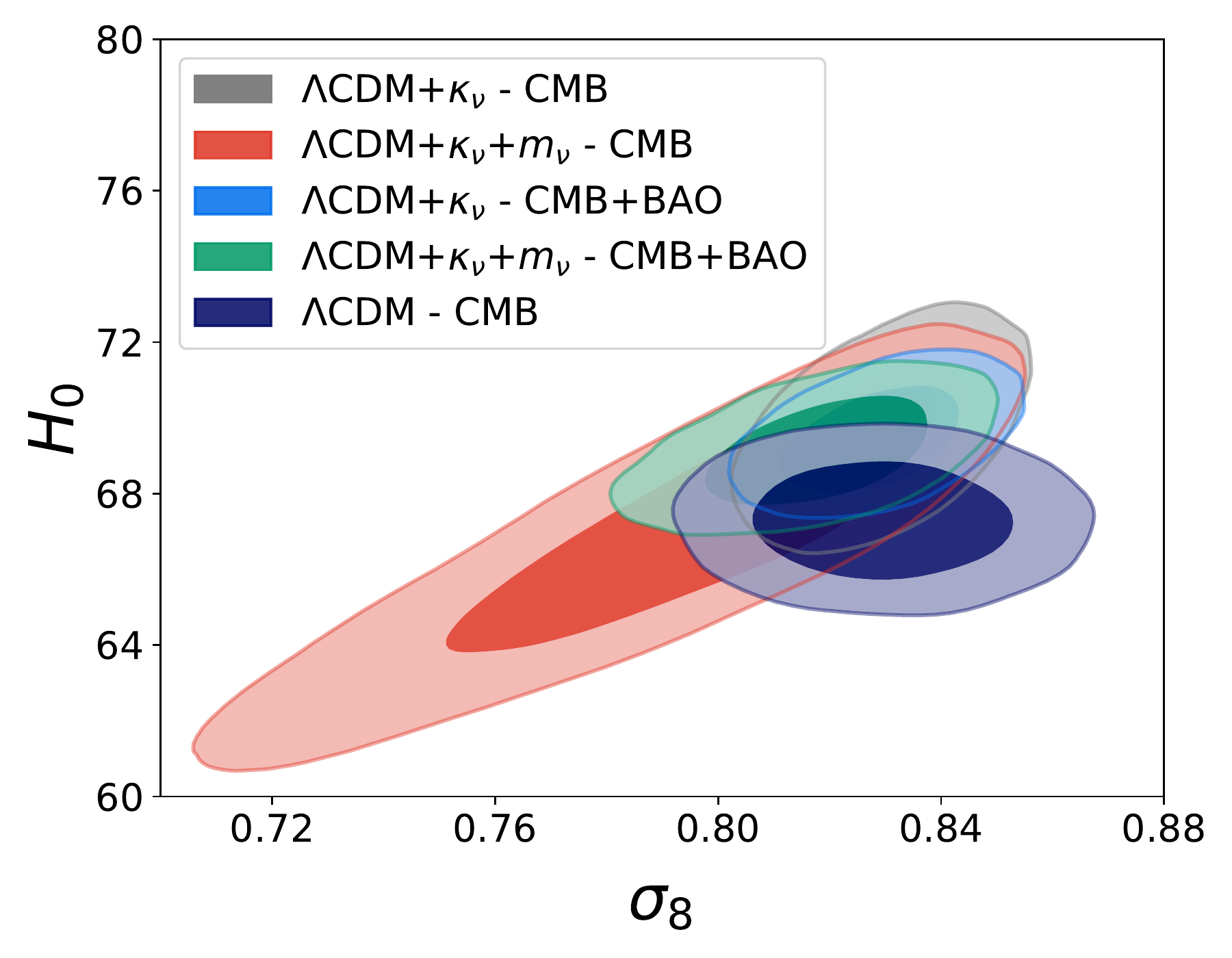}
\caption{\label{fig:s8_h0}
Constraints on $H_0$ and $\sigma_8$ 
in the \lcdmk\ and \lcdmkm\ models,
using CMB data alone and in combination with BAO,
in comparison with the result from the base \lcdm\ model (CMB only).
We also plot the local measurement of $H_0$ \cite{Riess:2016jrr}
in the corresponding panel.
Areas represent $1\sigma$ and $2\sigma$ credible regions.
}
\end{figure*}

We have also studied the impact of a mixed neutrino statistics on the $H_0$ and $\sigma_8$ tensions mentioned in the introduction.
In fig.~\ref{fig:s8_h0}, we show the comparison of the constraints on $H_0$ and $\sigma_8$
from the standard \lcdm\ model and
the extended models with free neutrino statistics.
As it is clear from the 1D posterior distributions,
the extended models produce a higher $H_0$ but similar values of
$\sigma_8$ with respect to the standard \lcdm\ model predictions.
Since the value of the Hubble parameter is larger,
the tension between CMB-based $H_0$ estimates and
its determinations from local measurements \cite{Riess:2016jrr}
is alleviated. Unfortunately, the 2D plot in the lower panel of fig.~\ref{fig:s8_h0}
shows very well that there is a strong correlation between $\sigma_8$
and $H_0$, telling us that if $H_0$ increases,
$\sigma_8$ cannot be significantly smaller
than its value in the \lcdm\ model.
This result indicates that a modified neutrino statistics alone
would not be enough to solve both the $H_0$ and $\sigma_8$ tensions simultaneously,
regardless of the assumptions on neutrino masses.

\begin{table}
\centering
\begin{tabular}{c|ccc}
\multicolumn{4}{c}{CMB only}\\
\hline
& 
\lcdmk & 
\lcdmm &
\lcdmkm \\
\hline
\lcdm & 0.01 & 0.51 & 1.59 \\
\lcdmk & - & 0.50 & 1.57 \\
\lcdmm & - & - & 1.07 \\
\end{tabular}
\caption{\label{tab:comparisonCMB}
Natural logarithm of the Bayes factors, $|\ln B_{ij}|$,
comparing the models adopted in this work,
using CMB data only.
Positive numbers indicate a preference for the model in the corresponding row.
The absolute errors on the various $|\ln B_{ij}|$ are all close to $0.3$.
}
\end{table}

\begin{table}
\centering
\begin{tabular}{c|ccc}
\multicolumn{4}{c}{CMB + BAO}\\
\hline
& 
\lcdmk & 
\lcdmm &
\lcdmkm \\
\hline
\lcdm & -0.42 & 1.30 & 2.02 \\
\lcdmk & - & 1.72 & 2.44 \\
\lcdmm & - & - & 0.72 \\
\end{tabular}
\caption{\label{tab:comparisonBAO}
The same as table~\ref{tab:comparisonCMB},
but using CMB+BAO data.}
\end{table}

Having performed the analyses with the \texttt{MultiNest} algorithm,
we can use the tools of
Bayesian model comparison to determine
if the introduction of the new additional degrees of freedom in the theoretical model
is motivated by the improvement of the data fit.
We briefly summarize the theory of Bayesian model comparison in appendix~\ref{sec:bayesian}.
We compare the standard \lcdm\ model with its three extensions (\lcdmk, \lcdmm\ and \lcdmkm),
using either CMB data only or the combination CMB+BAO.
As we can see from table~\ref{tab:comparisonCMB},
where we report the Bayes factors obtained considering
CMB data alone,
the simple \lcdm\ model remains the preferred one,
although the Bayes factors tell us that it is statistically
equivalent to the extended models
\lcdmk\ and \lcdmm\ according to the Jeffreys' scale
(see appendix~\ref{sec:bayesian}).
The most complex \lcdmkm\ model
is always weakly disfavoured with respect
to any of the other three choices.
The situation changes a bit when the BAO data are also considered,
see table~\ref{tab:comparisonBAO}:
in this case, the \lcdmk\ produces a smaller
Bayesian evidence than the \lcdm\ model,
that however is not significant according to the Jeffreys' scale.
If one has to decide which of the one-parameter extensions of
the \lcdm\ model is better,
the Bayes factors tell us that leaving $\kanu$ free is
a (weakly) favoured choice over leaving $m_\nu$ free.
The \lcdmkm\ model
is almost moderately disfavoured with respect to
the \lcdm\ one.

In a similar way, we can use the Savage-Dickey density ratio (SDDR)
approximation for the Bayes factor
to compute how much a purely bosonic neutrino
is disfavoured with respect to a purely fermionic one.
In our case, the SDDR approximation works because
the simpler model
$\mathcal{M}_{\kanu=\pm1} \equiv
\left.\Lambda\text{CDM}+\kanu \right|_{\kanu=\pm1}$
with purely fermionic ($+$) or bosonic ($-$)
neutrinos is nested within the extended model
$\mathcal{M}_{\kanu} \equiv \Lambda\text{CDM} + \text{(free) }\kanu$.
Given that \cite{Trotta:2008qt}
\begin{equation}\label{eq:sddr1}
 B_{\kanu=\pm1, \text{(free) }\kanu} =
 \left.\frac{p(\kanu|d,\mathcal{M}_{\kanu})}{p(\kanu|\mathcal{M}_{\kanu})}
 \right|_{\kanu=\pm1}\,,
\end{equation}
where $p(\kanu|d,\mathcal{M}_{\kanu})$ and $p(\kanu|\mathcal{M}_{\kanu})$
are the marginalized posterior and the prior
for \kanu\ in the extended model $\mathcal{M}_{\kanu}$,
which must be evaluated at the proper $\kanu=\pm1$.
Using the above definition,
the Bayes factor for the purely fermionic versus the purely bosonic
case can be written as:
\begin{equation}\label{eq:sddr2}
 B_{FD, BE} =
 \frac
 {p(\kanu|d,\mathcal{M}_{\kanu})|_{\kanu=+1}}
 {p(\kanu|d,\mathcal{M}_{\kanu})|_{\kanu=-1}}\,,
\end{equation}
so that we can compute the Bayes factor directly
from the posterior distribution functions depicted in fig.~\ref{fig:kappanu}.
Remembering that a positive $\ln B_{FD, BE}$ corresponds
to a preference for fermionic neutrinos,
we find that
\begin{equation}\label{eq:fd_vs_be_fixedmass}
\ln B_{FD, BE}\simeq1.0\; (1.2)
\qquad
\text{assuming }m_\nu=0.02\text{ eV},
\end{equation}
when considering CMB (CMB+BAO) data and a fixed neutrino mass.
If we repeat the exercise leaving also $m_\nu$ free to vary,
these numbers change to
\begin{equation}\label{eq:fd_vs_be_freemass}
\ln B_{FD, BE}\simeq1.6\; (0.5)
\qquad
\text{assuming free }m_\nu,
\end{equation}
given CMB (CMB+BAO) data.
These results indicate that cosmology can only weakly disfavour
purely bosonic neutrinos over the purely fermionic ones.

\section{Bounds from Primordial Nucleosynthesis}
\label{sec:bbn}

The implications on BBN of modified statistics for neutrinos were considered in the pioneering papers \cite{Cucurull:1995bx,Dolgov:2005qi,Dolgov:2005mi}. In particular, it was found that smaller values of $\kanu$ lead to a decrease both in the primordial mass fraction of helium-4, $Y_p$,  and in the produced $^7$Li abundance, while the amount of deuterium was increased  \cite{Dolgov:2005mi}.
At that time, primordial nucleosynthesis did not exclude a pure bosonic nature of neutrinos,
but a mixed neutrino statistics
went in the direction of improving the agreement between
the predicted value of the baryon asymmetry $\eta_{10} \equiv 10^{10}n_B/n_\gamma$
and its determination by CMB observations,
which actually constrain the baryon energy density $\omega_b$,
related to $\eta_{10}$ by
\begin{equation}
\omega_b \equiv \Omega_b h^2 = \frac{1 - 0.007125 Y_p}{273.279} \left( \frac{T^0_\gamma}{2.7255\,\mathrm{K}} \right)^3 \eta_{10}\,,
\end{equation}
where $T^0_\gamma$ is the CMB temperature today.

Neutrino statistics, through the form of modified neutrino distribution functions, enter BBN in two main ways:
\begin{itemize}
\item
the total energy density of neutrino flavours, $\rho_\nu$, contributes to the energy density of radiation, which in turn determines the Hubble expansion rate;
\item
the neutrino distribution functions directly appear in the weak rates that determine the neutron-proton chemical equilibrium.%
\end{itemize}
Both effects can significantly affect the final abundances of primordial nuclides from BBN.

We have used a modified version of the {\sc PArthENoPE} 2.0 code \cite{Pisanti:2007hk,Consiglio:2017pot} to find the theoretical prediction of primordial abundances when neutrinos present mixed statistics.
On one hand, we have accounted for the modification of the neutrino energy density in the non-standard case of $\kanu\neq 1$.
Since neutrinos are always relativistic at BBN, we have included the corresponding multiplicative factor on $\rho_\nu$, as depicted in Fig.~\ref{fig:rhonu}.
On the other hand, implementing a modified statistics in the calculation of the weak rates is a more complex problem, 
due to the fact that the operation requires a high degree of accuracy, which forces us to take into account radiative corrections to neutron-proton exchange rates 
(see \cite{Iocco:2008va} for a review on BBN physics).
In order to make a reliable estimate of the primordial abundances, 
we used in the modified {\sc PArthENoPE} the Born weak rates.
Additionally, we use the difference between the calculation including full radiative corrections and the one using Born rates,
both computed for $\kanu = 1$ and for the same values of the other parameters,
to revise the result and compensate for the lack of radiative corrections when using the mixed statistics.
 
\begin{figure*}[t]
\centering
\includegraphics[width=0.85\textwidth]{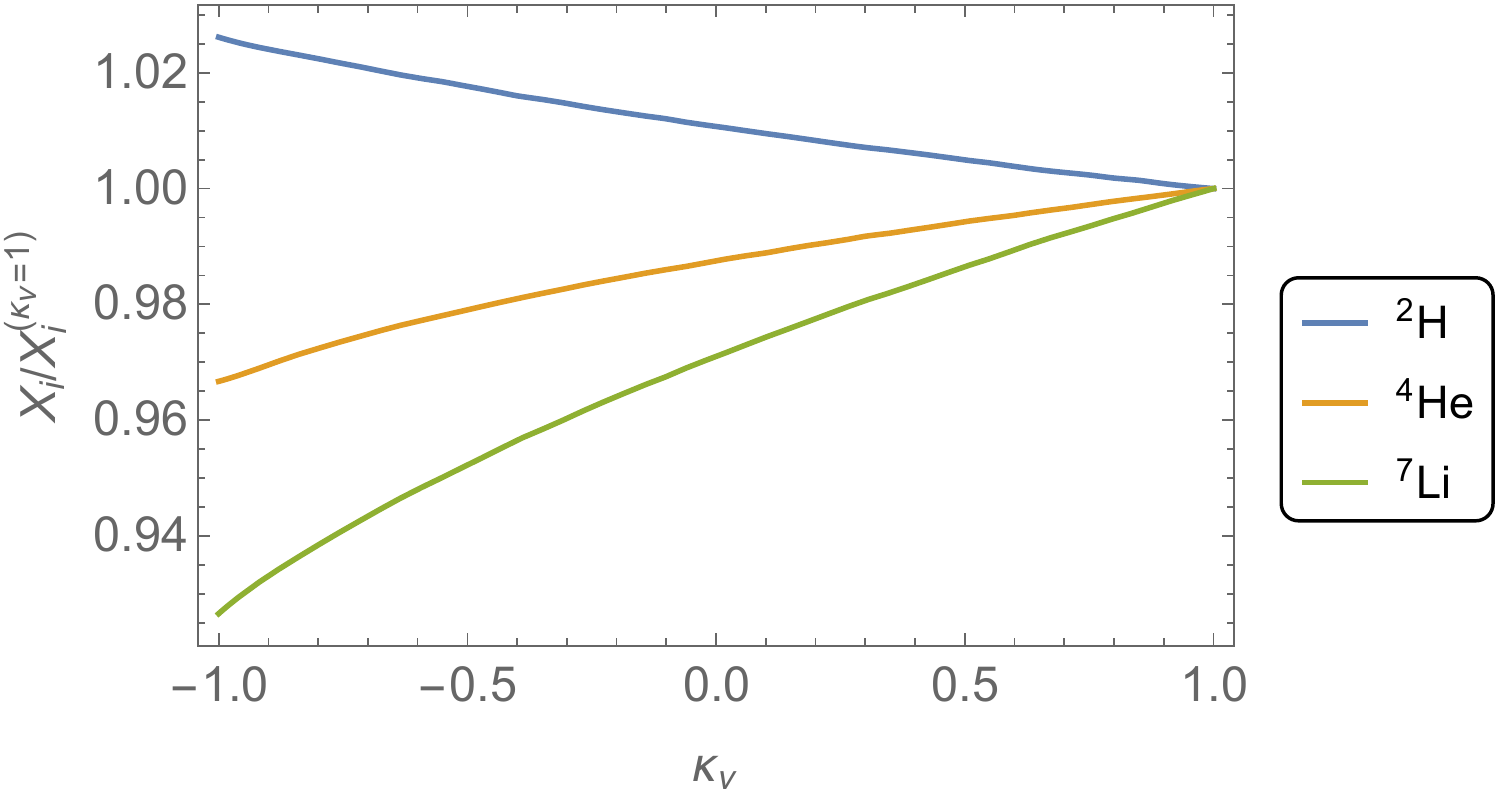}
\caption{\label{fig:abund}
Relative change of the primordial abundances, $X_i/X^{(\kanu=1)}_i$, as a function of $\kanu$, for $\eta = 6.094\cdot 10^{-10}$.}
\end{figure*}

Our BBN results for a particular value of the baryon asymmetry ($\eta_{10} = 6.094$) are shown in fig.~\ref{fig:abund}, where the relative change of the primordial abundances is plotted as a function of the effective Fermi-Bose parameter. The behaviour is in fair agreement with the results in figure 2 (lower panel) of \cite{Dolgov:2005mi}: the relative change in the primordial abundances is of the order of a few per cent, negative for 
$^4$He and $^7$Li and slightly positive for $^2$H.
Interestingly, we can note from the reduction of the produced $^4$He for smaller values of $\kanu$ that the direct effect of the neutrino distribution on the weak rates is more important
than the increased neutrino contribution to the energy density.

For our BBN analysis, we consider the following experimental determinations of the primordial abundances of deuterium \cite{Cooke:2017cwo},
helium-4 \cite{Aver:2015iza} and lithium-7 \cite{Sbordone:2010zi}:
\begin{eqnarray}
^{2}\mathrm{H/H} &=& \left( 2.527\pm 0.030 \right) \cdot 10^{-5}\,, \\
\mathrm{Y_p} &=& 0.2449\pm 0.0040\,, \\
^{7}\mathrm{Li/H} &=& \left( 1.58\pm 0.31 \right) \cdot 10^{-10}\,.
\end{eqnarray}
The theoretical values of the primordial abundances are compared with these measurements 
by defining the following $\chi^2$-functions:
\begin{equation}\label{eq:bbn_chi2}
\chi^2_i \left( \kanu, \eta_{10} \right) = \frac{\left( X_i^{\rm th}\left( \kanu, \eta_{10} \right) - X_i^{\rm exp} \right)^2}{\sigma^2_{i,{\rm th}}+\sigma^2_{i,{\rm exp}}}\,,
\end{equation}
where $i =\{\, ^{2}\mathrm{H/H}\, ,\, Y_p\, , ^{7}\mathrm{Li/H}\}$, $\sigma_{i,{\rm exp}}$ are the experimental errors as given above and $\sigma_{i,{\rm th}}$ is the uncertainty due to propagation of nuclear process rates: $\sigma_{^{2}\mathrm{H/H},{\rm th}}=0.05\cdot 10^{-5}$, \mbox{$\sigma_{Y_p,{\rm th}}=0.0003$,} and $\sigma_{^{7}\mathrm{Li/H},{\rm th}}=0.26\cdot 10^{-10}$ \cite{Coc:2015bhi}.
 
\begin{figure*}[t]
\centering
\includegraphics[width=0.85\textwidth]{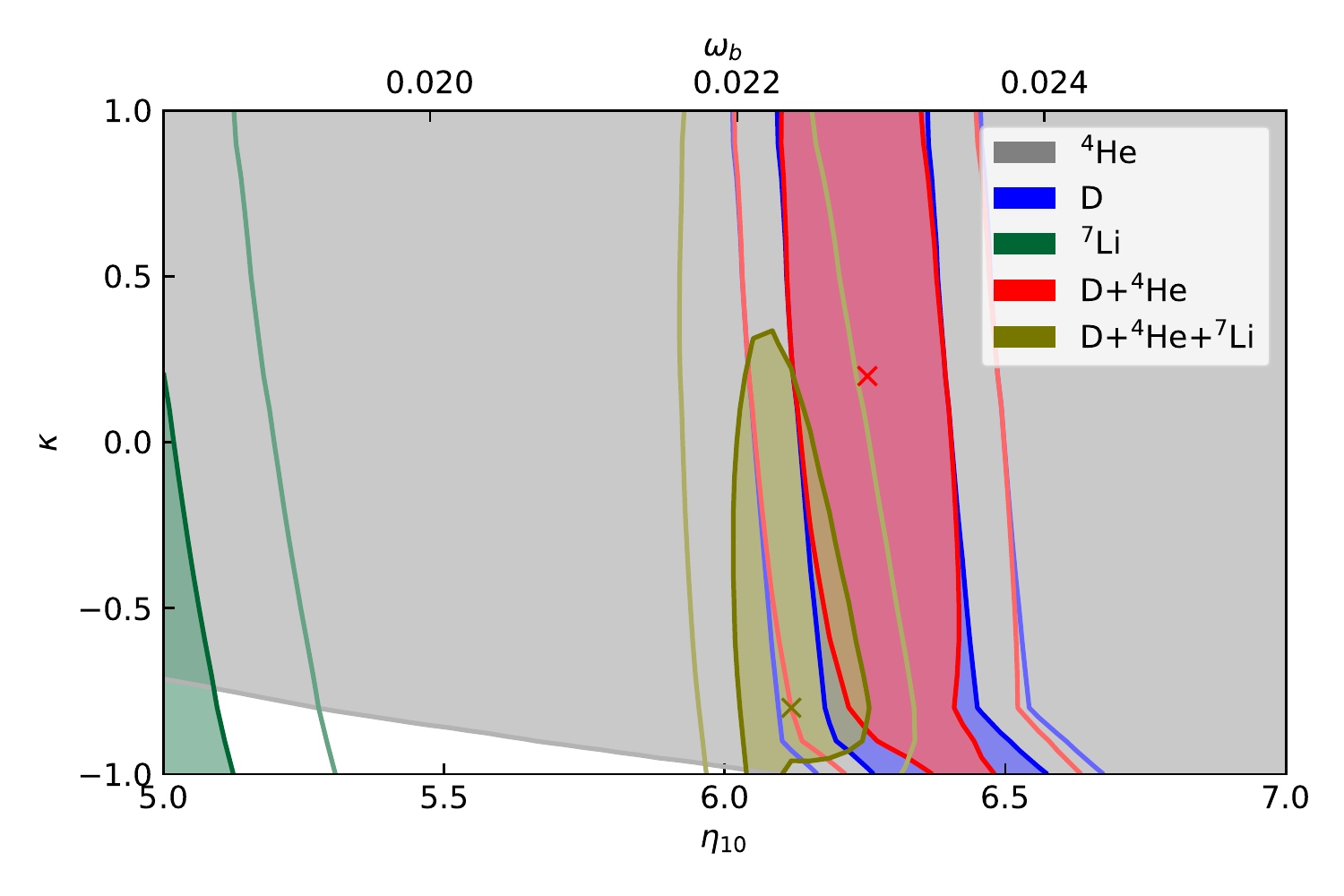}
\caption{\label{fig:noprior}
1- and 2-$\sigma$ contours in the plane ($\eta_{10}$, $\kanu$) from our BBN analysis. Areas in colour indicate the 1-$\sigma$ region and the corresponding line the 2-$\sigma$ one. In addition to the single abundance cases, we report the results for the combined ones, D$+^4$He and D$+^4$He$+^7$Li. Crosses mark the best-fit values for the combined cases.}
\end{figure*}

The results of our analysis are shown in Fig.~\ref{fig:noprior} as 1- and 2-$\sigma$ contours in the plane ($\eta_{10}$, $\kanu$), separately for each of the abundances and for the combined cases, D$+^4$He and D$+^4$He$+^7$Li.
Some comments are necessary.
While deuterium prefers a restricted region in $\eta_{10}$ (as it is a well known ``baryometer"), the 1-$\sigma$ region from $^4$He extends to almost all the plotted range and, at 2-$\sigma$, the whole plane is allowed.
At the same time, no evident indication for a preferred $\kanu$ value can be established, with the exception of the bottom left corner that is excluded by $^4$He, but only with 1-$\sigma$ significance and for values of $\omega_b$ in tension with CMB estimations.
The lithium problem is evident in the displacement of the 1-$\sigma$ lithium region to the extreme left (and almost out of the considered region).
The combined analysis without lithium essentially coincides with the area fixed by deuterium, with a mild preference for $\kanu\neq -1$.
Instead, when $^7$Li is included, the preference of this nuclide for a lower value of $\eta_{10}$ tends to favour the ``bosonic" character of neutrinos.
Again, this is a consequence of the existence of the $^7$Li problem,
which is only slightly alleviated but cannot be solved with 
non fermionic neutrinos.

\begin{figure*}[t]
\centering
\includegraphics[width=0.85\textwidth]{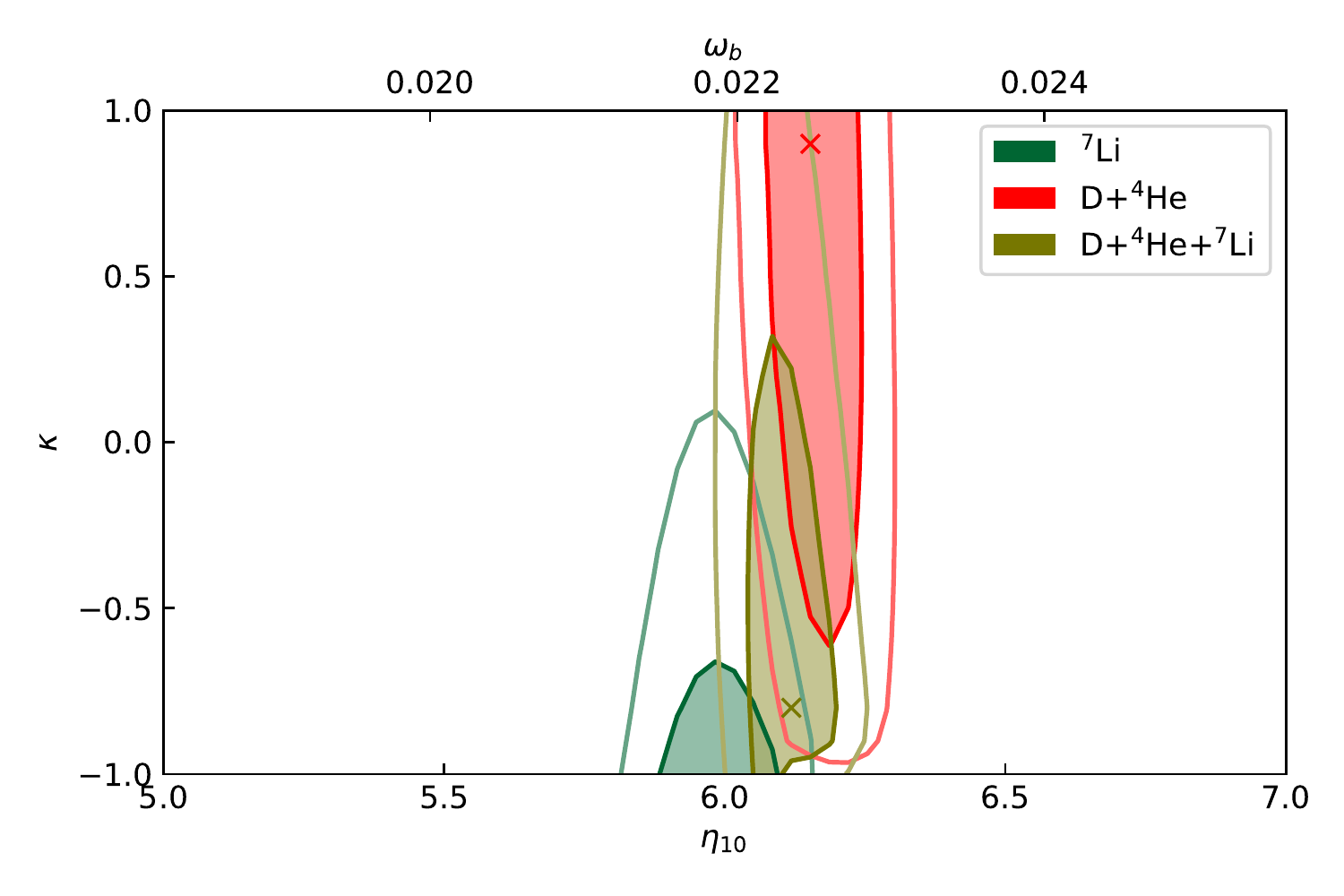}
\caption{\label{fig:prior}
Same as fig.~\ref{fig:noprior} including a prior on $\omega_b$ from CMB (see text).}
\end{figure*}

Finally, we want to investigate if a joint analysis of BBN, CMB and late-time cosmological observables could provide a stronger bound on neutrino statistics. 
Here we just consider the inclusion of a Gaussian prior on $\omega_b$ from CMB on the BBN analysis.
The prior, $\omega_b = 0.02230\pm 0.00027$, was derived accounting properly for neutrino statistics, i.e., it is the result of a marginalization over all the parameters except $\kanu$ of the full posteriors obtained using the extended model $\Lambda$CDM+$\kanu$+$m_\nu$ and only CMB data, discussed in section~\ref{sec:cosmo}.
The results are shown in Fig.~\ref{fig:prior}, where one can see that the prior helps in reducing the allowed $\eta_{10}$ region and, as a consequence, to improve the significance of the allowed region in $\kanu$.
In both cases, with or without prior, it is evident that D$+^4$He and $^7$Li prefer {\em more fermionic} and {\em more bosonic} neutrinos, respectively.
However, as we have already mentioned, the preference of $^7$Li for negative values of $\kanu$ should be taken with care, since it is 
an artefact of the disagreement between the experimental value and the prediction for the lithium abundance (i.e.\ the lithium problem).

As we did in the previous section,
we computed the Bayes factor of purely FD versus purely BE neutrinos
using the SDDR approximation, when considering BBN data alone
and in combination with the CMB prior on $\omega_b$.
To do so, we used \texttt{MultiNest} to perform a Bayesian exploration
of the two-parameter space $(\kanu, \eta_{10})$
and build the marginalized posterior distributions for \kanu\
given the BBN likelihood
$\mathcal{L}(\kanu, \eta_{10})
= \Pi_i \mathcal{L}_i(\kanu, \eta_{10})$,
where $i =\{\, ^{2}\mathrm{H/H}\, ,\, Y_p\, , ^{7}\mathrm{Li/H}\}$, $\mathcal{L}_i=\exp(-\chi_i^2/2)$
and the $\chi_i^2$ are the ones in eq.~\eqref{eq:bbn_chi2}.
Using again eq.~\eqref{eq:sddr2},
we find that the preference for fermionic neutrinos
may vary from inconclusive
($\ln B_{FD, BE}\simeq0.8$,
using D$+^4$He measurements alone)
to moderately significant
($\ln B_{FD, BE}\simeq3$,
when the $\omega_b$ constraint from CMB is also included as a prior),
depending on the data one considers.

\section{Summary and conclusions}
\label{sec:summary}

In this paper we have considered the cosmological implications of a scenario where neutrinos violate the Pauli principle and do not follow the spin-statistics relation. This assumption, which immediately arises theoretical problems that are difficult to solve, is analysed with a phenomenological approach via the
introduction of the so-called Fermi-Bose parameter $\kanu$, that characterizes mixed neutrino statistics. Current bounds on $\kanu$ from laboratory experiments are restricted to the observation of two-neutrino double beta decay, which only disfavours the region close to the purely bosonic case. Our aim was to check whether the analysis of cosmological observables could provide better bounds on neutrino statistics.

Neutrinos that do not necessarily follow a Fermi-Dirac distribution in equilibrium would alter the thermal history of the Universe, in a way that mimics the effect of a change in $\Neff$ at early times but is equivalent to an enhanced neutrino mass at late times. The modified neutrino energy distributions also affect the weak rates that fix the primordial production of nuclei during BBN. Our study shows that, despite the availability of very precise cosmological data, only very weak bounds are obtained on neutrino statistics. Using either CMB data alone or in combination with BAO results, we find that the region of $\kanu$ close to the purely bosonic case is weakly disfavoured, but only at less that $2\sigma$, with lower bounds at the level of $\kanu > -0.18$ to $\kanu > -0.06$. The same applies to the BBN analysis: only the smallest values of $\kanu$ are disfavoured, but at very low significance, once a prior on the baryon asymmetry from CMB is taken into account. We also find that the possibility of mixed neutrino statistics does not provide a way to solve current cosmological tensions. While we observe that changing neutrino statistics can bring the CMB-derived Hubble constant to a better agreement with the value obtained from the local measurement, it worsens the $\sigma_8$ tension. On the other hand, $\kanu\neq 1$ could alleviate but not solve the so-called $^7$Li problem of BBN. 

We conclude that cosmology can not provide at the moment a restrictive limit on neutrino statistics.
However, the next generation of cosmological surveys will be able to determine the neutrino energy density with exquisite precision
(see e.g.\ the recent forecast \cite{Sprenger:2018tdb})
and may improve significantly the
lower limit on $\kanu$,
complementary to the measurement of two-neutrino double beta decay.

\acknowledgments
P.F.d.S., S.G.\ and S.P.\  were supported by the Spanish grants FPA2017-85216-P and SEV-2014-0398 (MINECO), PROMETEOII/2014/084 (Generalitat Valenciana) and FPU13/03729 (MECD).
O.P.\ was supported by INFN under
Iniziativa Specifica TASP.
N.T.\ would like to thank Department of Physics and Astronomy, University of Padova, for the hospitality.
We thank Alexander Barabash for useful comments on double beta decay experiments.

\appendix
\section{Bayesian model comparison}
\label{sec:bayesian}
In this section we briefly report the basics of model comparison
in a Bayesian statistic framework.
For a complete review on the subject,
see for example ref.~\cite{Trotta:2008qt}.

In the following, we will indicate with $d$ the data under consideration
and with $\theta$ the parameters that describe a theoretical model $\mathcal{M}_i$.
Let us start defining the marginal likelihood or Bayesian evidence $Z_i$ of a model $\mathcal{M}_i$ as
\begin{equation}\label{eq:bayesevidence}
 Z_i =
 p(d|\mathcal{M}_i) =
 \int_{\Omega_{\mathcal{M}_i}}
 p(d|\theta, \mathcal{M}_i) \,
 p(\theta|\mathcal{M}_i)\,
 d\theta~,
\end{equation}
where $\Omega_{\mathcal{M}_i}$ is the entire parameter space allowed for the model,
$p(d|\theta, \mathcal{M}_i)$ is the 
likelihood of the data $d$ given the parameters $\theta$ from the model $\mathcal{M}_i$,
and $p(\theta|\mathcal{M}_i)$ is the prior
of the parameter combination $\theta$ within the model.
The Bayesian evidence is the central quantity if one wants to 
perform model comparison, because it can be related to the
posterior probability $P(\mathcal{M}_i|d)$ through the Bayes' theorem,
\begin{equation}\label{eq:modelpostprob}
 P(\mathcal{M}_i|d)  \propto P(\mathcal{M}_i)\, p(d|\mathcal{M}_i)~,
\end{equation}
where $P(\mathcal{M}_i)$ is the prior probability associated
to the model $\mathcal{M}_i$ under consideration.
If there are no theoretical reasons for which a model $\mathcal{M}_j$
should be preferred over the other models $\mathcal{M}_i$,
the quantities $P(\mathcal{M}_i)$ are the same.
If one wants to see whether the data $d$ prefer a model
$\mathcal{M}_j$ over $\mathcal{M}_k$,
the quantity to be computed is the ratio of the posterior probabilities:
\begin{equation}\label{eq:modelcomparison}
 \frac{p(\mathcal{M}_j|d)}{p(\mathcal{M}_k|d)}
 =
 B_{jk}
 \frac{p(\mathcal{M}_j)}{p(\mathcal{M}_k)}~,
\end{equation}
where the quantity $B_{jk}$ is the ratio of the evidences of the two models, also called Bayes factor,
\begin{equation}\label{eq:bayesfactor}
 B_{jk} = \frac{Z_j}{Z_k} \quad\Longrightarrow \ln B_{jk} = \ln Z_j -\ln Z_k~.
\end{equation}
A Bayes factor larger (smaller) than $1$ would imply a preference
for $\mathcal{M}_j$ over $\mathcal{M}_k$
($\mathcal{M}_k$ over $\mathcal{M}_j$),
given the data $d$.
The magnitude of $B_{jk}$ indicates 
how strong is the preference for one of the competing models
over the other
and it is usually determined by means of the Jeffreys' scale (see
Ref.~\cite{Trotta:2008qt} and references therein).
In particular,
$|\ln B_{jk}| \lesssim 1$ means inconclusive,
$1\lesssim |\ln B_{jk}| \lesssim 2.5$ means weak
and $2.5\lesssim |\ln B_{jk}| \lesssim 5$ means moderate statistical evidence.
A strong preference for one of the models is found when
$|\ln B_{jk}| \gtrsim 5$.

\bibliography{ref2}
   
\end{document}